\documentclass[aps,prb,showpacs,twocolumn,superscriptaddress]{revtex4}
\usepackage{graphicx}
\usepackage{bm}
\usepackage{amsmath}

\def\be{\begin{equation}}       \def\ee{\end{equation}}
\def\bea{\begin{eqnarray}}      \def\eea{\end{eqnarray}}

\begin{document}

\title{Absence of vertex correction for the spin Hall effect in
p-type semiconductors}
\author{Shuichi Murakami}
\email[Electronic address: ]{murakami@appi.t.u-tokyo.ac.jp}
\affiliation{Department of Applied Physics, University of Tokyo,
Hongo, Bunkyo-ku, Tokyo 113-8656, Japan}

\begin{abstract}
We calculate an effect of spinless impurities on the spin Hall effect 
of the Luttinger model representing p-type semiconductors. 
The self-energy in the Born approximation becomes diagonal in 
the helicity basis and its value is independent of the 
wavenumber or helicity. The vertex correction in the ladder approximation
vanishes identically, in sharp contrast with the Rashba model.
This implies that in the 
clean limit the spin Hall conductivity reproduces the value of the
intrinsic spin Hall conductivity calculated in earlier papers.
\end{abstract}

\maketitle

Understanding the dynamics of spins is a long-standing subject in 
semiconductor physics.
Though there have been lots of experimental and theoretical studies toward
this goal, there remain many obstacles to overcome before we can 
manipulate spins at our disposal in semiconductor 
spintronics devices \cite{wolf2001,awschalombook}.
One of the obstacles is efficient spin injection into semiconductors,
and there have been many atempts for it. For example,
spin injection from ferromagnetic metals has low efficiency,
because of the band structure mismatch between semiconductors and 
metals, in particular by the conductivity mismatch \cite{schmidt2000}.
On the other hand, while spin injection from 
ferromagnetic semiconductors such as (Ga,Mn)As \cite{ohno1998}
provides relatively high efficiency \cite{ohno1999b},
the Curie temperatures of such ferromagnetic semiconductors 
still remain lower than room temperature.
Thus the spin injection, which constitutes the very first step 
for spintronic devices, is still under intensive investigation.

In such circumstances, recent theoretical proposals of an intrinsic 
spin Hall effect \cite{murakami2003,sinova2004} 
have attracted much attention. 
In these works, it is theoretically predicted that an external 
electric field induces a dissipationless spin current in semiconductors.
This intrinsic spin Hall effect
is predicted for the Luttinger model representing 
bulk p-type semiconductors \cite{murakami2003},
and for the Rashba model representing 
two-dimensional n-type semiconductors in heterostructure 
\cite{sinova2004}. This intrinsic spin Hall effect is
caused by the Berry phase \cite{berry1984} in the momentum space.
Berry-phase structure in momentum space is comprised in
the electronic Bloch bands \cite{fang2003,yao2004,murakami2003b}, 
and endows the carriers with 
an  anomalous velocity.
Due to the spin-orbit coupling, the
carriers will then have spin-dependent 
trajectories in the presence of an external electric field, 
and it amounts to the intrinsic spin Hall effect.
In bulk
p-type semiconductors, 
the valence band has a spin 
splitting, giving rise to spin Hall effect by hole-doping
as proposed by the author and his collaborators
\cite{murakami2003},
whereas in bulk n-type semiconductors, 
the conduction band is doubly degenerate 
(if the Dresselhaus term is neglected),
and the spin Hall effect does not emerge.
Instead, by making the heterostructure from the n-type semiconductor,
the Rashba spin splitting \cite{rashba1960,bychkov1984}
arises, and it gives rise to the spin Hall effect 
proposed by Sinova et al. \cite{sinova2004}
Although an unambiguous experimental detection remains 
to be done, many theoretical studies have appeared
since the proposals of the intrinsic spin Hall effect
\cite{murakami2003c,culcer2003,burkov2003,hu2004,xiong2003,inoue2004,%
hu2003,bernevig2003,rashba2003,shen2003,sinitsyn2003,shen2004,%
schliemann2003}.
We note that these are different from an extrinsic spin Hall effect
caused by spin-dependent scattering by impurities, 
which might be assumed as a main source of 
spin Hall effect in earlier papers
\cite{dyakonov1971,hirsch1999,zhang2000}.

In order to argue possibilities of experimental detection, 
theoretical investigation of disorder effects is 
highly desired.
In this paper, we mainly focus on 
the Luttinger model representing p-type semiconductors, 
\cite{luttinger1956},
and study disorder effects on the spin Hall effect.
We shall also discuss on the Rashba model.
For the Luttinger model,
the effect of self-energy 
broadening was calculated by Schliemann and Loss \cite{schliemann2003}, 
and no other papers on disorder
effect have appeared at present. 
In this paper, 
we study the effect of randomly
distributed spinless impurities with short-ranged potential, by calculating
both the self-energy correction in the Born approximation 
and the vertex correction in the ladder approximation.
The calculation itself is analogous to that by Inoue et al.\ for the 
Rashba model \cite{inoue2003,inoue2004}. 
The resulting self-energy is of a simple form, 
independent of the wavevector $\mathbf{k}$.
Remarkably, the vertex correction turns out to vanish identically, 
even away from the clean limit. Thus the spin Hall conductivity becomes
identical with that of the intrinsic one calculated in 
ref.\ \onlinecite{murakami2003}. We note that it is in sharp contrast with the 
Rashba model, where the vertex correction completely kills the spin 
Hall effect within the Born and ladder approximations in the clean limit
\cite{inoue2003,inoue2004}.

The Luttinger Hamiltonian \cite{luttinger1956} is written as
\begin{equation}
H_{0}(\mathbf{k})=\frac{\hbar^{2}}{2m}
\left((\gamma_{1}+\frac{5}{2}\gamma_{2})k^{2}-
2\gamma_{2}(\mathbf{k}\cdot \mathbf{S}
)^{2}\right) \label{Luttinger}
\end{equation}
where $\mathbf{S}=(S^x,\ S^y,\
S^z)$ are the spin-$3/2$ matrices, 
$\mathbf{k}=(k_x,\ k_y,\ k_z)$, and $k=|\mathbf{k}|$.
For simplicity, we have put $\gamma_{2}=\gamma_{3}$ in the original
Luttinger Hamiltonian.
In this Hamiltonian, a helicity defined 
by $\lambda=\mathbf{k}\cdot\mathbf{S}/k$ is a good quantum number
and can be used as a label for eigenstates.
The helicity $\lambda$ can take values $\pm\frac{3}{2},\ \pm\frac{1}{2}$;
$\lambda=\pm\frac{3}{2}$ correspond to the heavy-hole (HH) band and 
$\lambda=\pm\frac{1}{2}$ to the light-hole (LH) band.
Thus the energy $E_{\mathbf{k}\lambda}$ of the 
eigenstatestate with wavenumber $\mathbf{k}$ and helicity $\lambda$ is
expressed as
\begin{eqnarray}
E_{k,\pm\frac{3}{2}}=E_{k H}=
\frac{\gamma_{1}-2\gamma_{2}}{2m}\hbar^{2}k^{2}\\
E_{k,\pm\frac{1}{2}}=
E_{k L}=\frac{\gamma_{1}+2\gamma_{2}}{2m}\hbar^{2}k^{2},
\end{eqnarray}
which only depends on $k=|\mathbf{k}|$.
By a unitary transformation $U_{\mathbf{k}}=e^{i\theta S_y}
e^{i\phi S_{z}}$, where $\theta$, $\phi$ are spherical coordinates of 
$\mathbf{k}$, $(\hat{k}\cdot\mathbf{S})$ is diagonalized as
\begin{equation}
U_{\mathbf{k}}(\hat{k}\cdot\mathbf{S})
U^{\dagger}_{\mathbf{k}}=S_{z}=\text{diag}\left(\frac{3}{2},\frac{1}{2},
-\frac{1}{2},-\frac{3}{2}\right),
\end{equation}
where $\hat{k}=\mathbf{k}/k$, and 
the Hamiltonian is diagonalized as well;
\begin{equation}
U_{\mathbf{k}}H_{0}U^{\dagger}_{\mathbf{k}}
=\text{diag}\left(E_{kH},\ E_{kL},\ E_{kL},\ E_{kH}\right).
\end{equation}
Therefore, an
eigenstate $|\mathbf{k}\lambda\rangle$
of the Hamiltonian $H$ with helicity $\lambda$
is expressed as
\begin{equation}
|\mathbf{k}\lambda\rangle=U^{\dagger}_{\mathbf{k}}|\lambda\rangle.
\label{klambda}
\end{equation}
where
$|\lambda\rangle$ denote the eigenstate of $S_{z}$ with 
the eigenvalue $\lambda$.

We consider randomly distributed nonmagnetic 
impuries with short-ranged potential:
\begin{equation}
V(\mathbf{r})=V\sum_{i}\delta(\mathbf{r}-\mathbf{R}_{i}).
\end{equation}
We shall calculate the Green function $G=(z-H)^{-1}$, $H=H_{0}+V$ in 
the self-consistent Born approximation from the  
unperturbed Green function $G^{(0)}=(z-H_{0})^{-1}$.
$G^{(0)}$ is diagonal in the helicity basis,
with elements $g^{(0)}_{k\lambda}(z)=(z-E_{k\lambda})^{-1}$.
By taking the average $\langle \cdots
\rangle_{\text{AV}}$ over the impurity distribution,
the lowest-order term of the 
self-energy in the Born approximation is written as 
\begin{eqnarray}
&&\left\langle\langle \mathbf{k}\lambda|VG^{(0)}V|\mathbf{k}''\lambda''
\rangle\right\rangle_{\text{AV}}\nonumber \\
&&\ \ =\frac{nV^{2}}{L^{3}}\delta_{\mathbf{k}\mathbf{k}''}
\sum_{\mathbf{k}',\lambda'}\langle\mathbf{k}\lambda |
\mathbf{k}'\lambda'\rangle
\langle\mathbf{k}'\lambda'|
\mathbf{k}\lambda''\rangle 
g^{(0)}_{k'\lambda'},
\label{self-energy}
\end{eqnarray}
where $n=N/L^{3}$ is a number density of the impurities.

We shall show that this self-energy (\ref{self-energy}) is 
written as $\Sigma \delta_{\mathbf{k}\mathbf{k}''}\delta_{\lambda
\lambda''}$, where $\Sigma$ is a constant.
By substituting (\ref{klambda}) into (\ref{self-energy}), we
get
\begin{eqnarray}
&&\left\langle\langle \mathbf{k}\lambda|VG^{(0)}V|\mathbf{k}''\lambda''
\rangle\right\rangle_{\text{AV}}\nonumber \\
&&=\frac{nV^{2}}{L^{3}}\delta_{\mathbf{k}\mathbf{k}''}
\sum_{\mathbf{k}',\lambda'}\langle \lambda |
U_{\mathbf{k}}U^\dagger_{\mathbf{k}'}
|\lambda'\rangle g^{(0)}_{k'\lambda'}
\langle\lambda'|
U_{\mathbf{k}'}U^\dagger_{\mathbf{k}}
|\lambda''\rangle.
\end{eqnarray}
We note that the Green function $g^{(0)}_{k'\lambda'}$ 
is written as $a^{(0)}_{k'}+b^{(0)}_{k'}(S_{z}^{2})_{\lambda'\lambda'}$,
where 
$a^{(0)}_{k'}+\frac{1}{4}b^{(0)}_{k'}
=(z-E_{k'L})^{-1}$, $a^{(0)}_{k'}+\frac{9}{4}
b^{(0)}_{k'}=(z-E_{k'H})^{-1}$.
Here $a^{(0)}_{k'}$ and $b^{(0)}_{k'}$ depend only on $k'=|\mathbf{k'}|$ and 
not  on the direction of $\mathbf{k}'$.
The self-energy is then rewritten as
\begin{eqnarray}
&&\left\langle\langle \mathbf{k}\lambda|VG^{(0)}V|\mathbf{k}''\lambda''
\rangle
\right\rangle_{\text{AV}}\nonumber \\
&&=\frac{nV^{2}}{L^{3}}\delta_{\mathbf{k}\mathbf{k}''}
\sum_{\mathbf{k}'}\left\langle \lambda \left|
U_{\mathbf{k}}U^\dagger_{\mathbf{k}'}
\left( a^{(0)}_{k'}+b^{(0)}_{k'}S_{z}^{2}\right)
U_{\mathbf{k}'}U^\dagger_{\mathbf{k}}
\right|\lambda''\right\rangle\nonumber \\
&&=\frac{nV^{2}}{L^{3}}\delta_{\mathbf{k}\mathbf{k}''}
\sum_{\mathbf{k}'}\left\langle \lambda \left|
U_{\mathbf{k}}
\left( a^{(0)}_{k'}+b^{(0)}_{k'}(\hat{k}'\cdot\mathbf{S})^{2}\right)
U^\dagger_{\mathbf{k}}
\right|\lambda''\right\rangle.\ \ 
\end{eqnarray}
Among various terms in $(\hat{k}'\cdot\mathbf{S})^{2}$, the 
contributions from 
cross-terms such as $\hat{k}'_{x}\hat{k}'_{y}S_{x}S_{y}$ vanish by
the summation over $\hat{k}'$.
Likewise,
in a term $({\hat{k}'}_{x})^{2}S_{x}^{2}$, for example,
$({\hat{k}'}_{x})^{2}$ can be replaced by 
$\frac{1}{3}(\hat{k}')^{2}=\frac{1}{3}$.
Hence, $(\hat{k}'\cdot\mathbf{S})^{2}$ can
be replaced by $\frac{1}{3}\mathbf{S}^{2}=\frac{5}{4}$, and
we get 
\begin{eqnarray}
&&\left\langle\langle \mathbf{k}\lambda|VG^{(0)}(z)V|\mathbf{k}''\lambda''
\rangle \right\rangle_{\text{AV}}\nonumber \\
&&=\frac{nV^{2}}{L^{3}}\delta_{\mathbf{k}\mathbf{k}''}
\sum_{\mathbf{k}'}\left\langle\lambda \left|
U_{\mathbf{k}}
\left( a^{(0)}_{k'}+\frac{5}{4}b^{(0)}_{k'}\right)
U^\dagger_{\mathbf{k}}
\right|\lambda''\right\rangle\nonumber \\
&&=\frac{nV^{2}}{2L^{3}}\delta_{\mathbf{k}\mathbf{k}''}
\delta_{\lambda\lambda''}
\sum_{\mathbf{k}'}\left(
\frac{1}{z-E_{k'L}}
+\frac{1}{z-E_{k'H}}\right).
\end{eqnarray}
By a similar argument, the full self-energy in the self-consistent Born 
approximation is seen to be diagonal in $\mathbf{k}$ and $\lambda$:
\begin{eqnarray}
&&\left\langle\langle
\mathbf{k}\lambda |G|\mathbf{k}''\lambda''\rangle\right\rangle_{\text{AV}}
=\frac{1}{(g^{(0)}_{k\lambda})^{-1}-\Sigma}\delta_{\mathbf{k}\mathbf{k}''}
\delta_{\lambda\lambda''}\nonumber\\
&&\ \ \ 
=g_{k\lambda}\delta_{\mathbf{k}
\mathbf{k}''}\delta_{\lambda\lambda''},\\
&&g_{\mathbf{k}H}=g_{\mathbf{k},\frac{3}{2}}
=g_{\mathbf{k},-\frac{3}{2}},
\ \ \ 
g_{\mathbf{k}L}=g_{\mathbf{k},\frac{1}{2}}
=g_{\mathbf{k},-\frac{1}{2}},
\label{Sigma}\\
&& \Sigma=\frac{nV^{2}}{2L^{3}}\sum_{\mathbf{k}'}(g_{k'L}
+g_{k'H}),
\end{eqnarray}
where $g_{k\lambda}$ is a full Green function in the 
self-consistent Born approximation.
The self-energy has several interesting properties; (i) it is diagonal
in $\mathbf{k}$ and $\lambda$, and (ii) it is independent 
of $\mathbf{k}$ and of $\lambda$.
It guarantees an assumption in the paper by Schliemann and Loss
\cite{schliemann2003}.

Let us calculate a vertex correction for a charge current vertex within
the ladder approximation.
The charge current operator $J_{x}$ is written as
\begin{eqnarray}
J_{x}=\frac{e}{\hbar}\frac{\partial H}{\partial k_{x}}
=\frac{e\hbar}{m}\left\{
\left(\gamma_{1}+\frac{5}{2}\gamma_{2}\right)k_{x}
-\gamma_{2}\left\{S_{x},\ 
\mathbf{k}\cdot\mathbf{S}\right\}
\right\},
\end{eqnarray}
where 
$\{\ , \ \}$ is the anticommutator.
The lowest order term of the vertex correction for $J_{x}$
in the ladder approximation is given by
\begin{eqnarray}
&&\left\langle\langle \mathbf{k}\lambda|
VG(z)J_{x}G(z')V|
\mathbf{k}'\lambda' 
\rangle\right\rangle_{\text{AV}}
\nonumber \\
&&\ =
\frac{V^2}{L^{6}}
\sum_{\mathbf{k}_{1},\lambda_{1}}\sum_{\mathbf{k}_{2},\lambda_{2}}
\sum_{i}\left\langle e^{-i(\mathbf{k}-\mathbf{k}_{1})\cdot\mathbf{R}_{i}}
e^{i(\mathbf{k}'-\mathbf{k}_{2})\cdot\mathbf{R}_{i}}\right\rangle_{\text{AV}}
\nonumber \\
&&\ \ \times 
\langle\mathbf{k}\lambda|\mathbf{k}_{1}\lambda_{1}\rangle
\langle\mathbf{k}_{2}\lambda_{2}|\mathbf{k}'\lambda'\rangle
g_{k_{1}\lambda_{1}}(z)
g_{k_{2}\lambda_{2}}(z')\langle \mathbf{k}_{1}\lambda_{1}|
J_{x}|\mathbf{k}_{2}\lambda_{2}\rangle\nonumber \\
&&\ =
\frac{nV^{2}}{L^{3}}\delta_{\mathbf{k}\mathbf{k}'}
\sum_{\mathbf{k}_{1},\lambda_{1},\lambda_{2}}
\langle\mathbf{k}\lambda|\mathbf{k}_{1}\lambda_{1}\rangle
\langle\mathbf{k}_{1}\lambda_{2}|\mathbf{k}\lambda'\rangle\nonumber \\
&&\ \ \ \ \times
g_{k_{1}\lambda_{1}}(z)
g_{k_{1}\lambda_{2}}(z')\langle \mathbf{k}_{1}\lambda_{1}|
J_{x}|\mathbf{k}_{1}\lambda_{2}\rangle.
\end{eqnarray}
From eq.\ (\ref{Sigma}), 
$g_{\mathbf{k}\lambda}$ can be written 
as $g_{\mathbf{k}\lambda}=a_{k}+b_{k}(S_{z}^{2})_{\lambda\lambda}$.
This relation is useful in the calculation of the vertex correction as 
demonstrated below.
\begin{eqnarray}
&&\left\langle\langle \mathbf{k}\lambda|
VG(z)J_{x}G(z')V|
\mathbf{k}'\lambda 
\rangle\right\rangle_{\text{AV}}
\nonumber \\
&&\ =
\frac{nV^{2}}{L^{3}}\delta_{\mathbf{k}\mathbf{k}'}
\sum_{\mathbf{k}_{1}}
\langle \lambda|
U_{\mathbf{k}}U^{\dagger}_{\mathbf{k}_{1}}(a_{k_{1}}(z)+b_{k_{1}}(z)S_{z}^{2})
U_{\mathbf{k}_{1}}
\nonumber \\
&&\ \ \times J_{x}(\mathbf{k}_{1})
U^{\dagger}_{\mathbf{k}_{1}}
(a_{k_{1}}(z')+b_{k_{1}}(z')S_{z}^{2})
U_{\mathbf{k}_{1}}U^{\dagger}_{\mathbf{k}}|\lambda'\rangle
\nonumber \\
&&\ =
\frac{nV^{2}}{L^{3}}\delta_{\mathbf{k}\mathbf{k}'}
\sum_{\mathbf{k}_{1}}\langle \lambda|
U_{\mathbf{k}}\{a_{k_{1}}+b_{k_{1}}(\hat{k}_{1}\cdot \mathbf{S})^{2}\}
\nonumber \\
&&\ \ \times J_{x}(\mathbf{k}_{1})
\{a_{k_{1}}+b_{k_{1}}(\hat{k}_{1}\cdot\mathbf{S})^{2}\}
U^{\dagger}_{\mathbf{k}}|\lambda'\rangle.
\end{eqnarray}
Because the summand is an odd function of $\mathbf{k}_{1}$, 
this lowest order term of the vertex correction in the 
ladder approximation vanishes identically. 
Therefore, higher order terms in the ladder approximation vanish as well, 
and thus we conclude that the vertex correction 
for the current $J_{x}$ in the ladder approximation vanishes.
This cancellation is between the intermediate states with 
$\mathbf{k}_{1}$ and $-\mathbf{k}_{1}$, i.e., due to parity.
It is similar to the familiar example of vanishing 
vertex correction for the Fermi gas with spinless isotropic
impurities.
When we turn to the calculation of the spin 
Hall conductivity, we have to take into account 
the vertex correction in the charge current only.
Therefore, 
because of this vanishing vertex correction, 
the spin Hall conductivity in the clean
case reproduces the intrinsic value, which is 
obtained from a clean system without impurities
from the outset \cite{murakami2003,murakami2003c,note-definition}. 
In other words, the result by taking a limit 
$1/\tau\rightarrow 0$ before $\omega\rightarrow 0$ is identical with 
that by the reverse order of limits.
The Luttinger model is free from a problem of order of limits.

By inspection, we can generalize the above discussion;
for any inversion-symmetric models 
with $H(\mathbf{k})=H(-\mathbf{k})$, the vertex correction 
vanish identically for 
short-ranged scatterers. For example, even if we introduce the anisotropy in 
Luttinger parameters $\gamma_{2}\neq\gamma_{3}$, the vertex correction 
remains zero.
If the impurity potential is long-ranged, forward scattering is
preferred and the vertex correction 
no longer vanishes. A simple calculation similar to that
for short-ranged scatterers shows that the resulting spin Hall effect 
is enhanced by a factor of $\tau_{\text{tr}}^{\text{SH}}/\tau$, 
where $\tau_{{\text tr}}^{\text{SH}}$ represents an effective transport
lifetime for the spin Hall effect.

This is to be contrasted with the Rashba Hamiltonian studied by Inoue
et al.\ \cite{inoue2003,inoue2004} 
They studied the vertex correction within the 
ladder approximation for randomly distributed nonmagnetic impurities
with isotropic potential. 
The result is remarkable;
with the vertex correction, the spin Hall effect becomes zero in the 
clean limit \cite{inoue2003,inoue2004,raimondi2001}, instead of $e/8\pi$,
the universal value without impurities \cite{sinova2004}. 
In their paper, they calculated the charge-current vertex appearing
in the correlation function between the charge current and the spin 
current in the Kubo formalism. In the clean limit, the vertex correction for 
the charge current in the ladder approximation is $-1$ times the 
spin-dependent part of the charge current operator; therefore, 
the total charge-current
vertex becomes spin-independent, yielding a vanishing spin Hall conductivity.
This vanishing result by Inoue et al.\ corresponds to the order of
limits, $\omega\rightarrow 0$ before $1/\tau\rightarrow 0$, while
the calculation of the intrinsic universal value 
by Sinova et al.\ corresponds to $1/\tau
\rightarrow 0$ before $\omega \rightarrow 0$.
The former and the latter limits correspond to $L>l$ and $L<l$, respectively,
where $L$ is the system size and $l$ is the mean free path.
Nonetheless, it is not the end of the
story. Two numerical calculations including disorder, 
one based on Kubo formula 
\cite{nomura2004} and the other on scattering theory \cite{xiong2003}, 
indicate that  the spin Hall effect in the clean limit remains 
the universal value $e/8\pi$. It contradicts the analytical result, 
and we do not have any answer to it.
Here, we also note that 
Burkov \textit{et al.}\ \cite{burkov2003} formulated a 
theory of spin-charge coupled transport, and
discuss that the spin Hall conductivity vanishes in a 
diffusive regime. Thus, a comprehensive understanding 
of disorder effects for various strength of disorder
is still to be desired.

Though the Rashba model is different from the Luttinger model,
our calculation on the Luttinger model has some implications on 
the debate on the disorder effect in the Rashba model.
To interpret the vanishing spin Hall conductivity in the clean limit
\cite{inoue2004},
Inoue et al.\ discuss that diffuse scattering efficiently 
scrambles the precession of spins such that
no net spin Hall current remains.
This picture is too simplified, since the spin Hall conductivity is not
necessarily zero for general systems with spin-orbit coupling.
Indeed, in the Luttinger model the 
vertex correction vanishes as we have seen.
Furthermore, even for more general models with Rashba coupling,
the spin Hall conductivity including the vertex correction 
is not necessarily zero in the clean limit.
For example, instead of the simplest model of Rashba coupling
\begin{equation}
H=\frac{\hbar^{2}k^{2}}{2m}+\hbar\lambda
(k_{y}\sigma_{x}-k_{x}\sigma_{y}),
\label{rashba}
\end{equation}
let us take a model
\begin{equation}
H=\frac{\hbar^{2}k^{2}}{2m}+\hbar(\lambda+\lambda_{1}k^{2})
(k_{y}\sigma_{x}-k_{x}\sigma_{y}),
\end{equation}
where $\lambda$ and $\lambda_{1}$ are constants. 
An extra term $\lambda_{1}$ is added here;
this term should exist in general because 
it is allowed by symmetry.
The vertex correction is calculated in the similar 
procedure as in refs.\ \onlinecite{inoue2003,inoue2004}, and
we can calculate
the coefficient $\lambda'$ for the vertex correction.
After lengthy but straightforward calculation, we 
can see that the spin Hall conductivity is nonzero
in the clean limit, 
even if the vertex correction is included within the 
ladder approximation. 
Thus, for the simplest model of Rashba coupling, (\ref{rashba}),
the complete cancellation of the 
spin Hall effect \cite{inoue2003,inoue2004} seems 
merely accidental, and not a consequence of any symmetries.
This is also supported by the result with long-ranged scatterers, 
where the spin Hall conductivity including the 
vertex correction no longer vanishes \cite{inoue2004}.
To summarize, we can say that the 
spin Hall effect is not necessarily suppressed 
to zero by the vertex correction in general.

In conclusion, we consider an effect of
spinless impurities on the spin Hall effect in the Luttinger model.
We calculated the vertex correction for the 
charge current within the ladder approximation.
For short-ranged scatters, the vertex correction is zero, and the 
spin Hall effect reproduces the intrinsic value, calculated previously 
from the system without impurities.

We thank G.\ E.\ W.\ Bauer, 
D.\ Culcer, J.\ Inoue, A.\ H.\ MacDonald, 
N.\ Nagaosa, Q.\ Niu, K.\ Nomura, J.\ Sinova, and 
S.\ -C.\ Zhang for fruitful discussions and helpful comments.
This work is
supported by Grant-in-Aids from the Ministry of Education,
Culture, Sports, Science and Technology of Japan.

\end{document}